\documentclass[11pt]{article}
\usepackage{graphicx}
\usepackage{latexsym,graphicx}
\usepackage{float}
\usepackage{amsmath}
\usepackage{amsfonts}
\usepackage{amssymb}
\usepackage{hyperref}
\usepackage{epstopdf}
\usepackage{color}
\usepackage{cite}
\usepackage{orcidlink}
\DeclareGraphicsExtensions{.eps}
\usepackage[left=2.5cm,top=2.5cm,right=2.5cm,bottom=2.5cm]{geometry}
\catcode `\@=11
\catcode `\@=12
\begin{document}
	\begin{center}
		\large{\bf{Transit dark energy cosmological models in generalized matter-geometry coupling theory using a non-linear form of $f(R,T,L_{m})$ function}} \\\vspace{5mm}
		\normalsize{Dinesh Chandra Maurya\orcidlink{0000-0003-2578-9629}$^{1, a}$, Rashid Zia\orcidlink{0000-0003-2047-314X}$^{2, b}$}\\
\vspace{5mm}
\normalsize{$^{a}$Centre for Cosmology, Astrophysics and Space Science, GLA University, Mathura-281 406,
	Uttar Pradesh, India.}\\
\vspace{2mm}
\normalsize{$^{b}$Department of Mathematics, United University, Prayagraj-211 012, Uttar Pradesh, India,}\\
\vspace{2mm}
$^{1}$E-mail:dcmaurya563@gmail.com \\
\vspace{2mm}
$^{2}$E-mail:rashidzya@gmail.com \\
	\end{center}
	\vspace{5mm}
	\begin{abstract}
		We have investigated the cosmological consequences of the model in the recently developed gravity theory [Haghani and Harko, \textit{Eur. Phys. J. C} \textbf{81} (2021) 615.] using a non-linear form of the $f(R,T,L_{m})$ function and the latest observational datasets. For flat Friedman-Lema\^{\i}tre-Robertson-Walker (FLRW) spacetime and $f(R,T,L_{m})= \alpha\,R+\beta\,RT+\gamma\,RL_{m}-\eta$ with $\alpha$, $\beta$, $\gamma$, and $\eta$ as coupling constants, we have solved the modified field equations to get the Hubble function $H(z)$ in terms of $H_{0}$, $\Omega_{m0}$, $\Omega_{r0}$, $\Omega_{\eta}$, $\beta$, and $\gamma$. To ensure that the model is consistent with the physically observed universe, we constrained the model parameters using Monte Carlo Markov Chain (MCMC) analysis on joint datasets of cosmic chronometer and Pantheon samples. Using these approximated model parameter values, we investigated the universe's cosmic evolution history, including the deceleration parameter, effective equation of state, dark energy equation of state, total dark energy density parameters, universe age, and so on. In addition, to assess the physical acceptability and stability of the generated model, we conducted the Om diagnostic test, causality test, and energy conditions test.
	\end{abstract}
	\smallskip
	\vspace{5mm}
	\vspace{1cm}
	
	\section{Introduction}
	
	In the light of the observations of distant supernovae by “The Supernova Cosmology Project” \cite{ref1,ref2} and “The High-Z Supernova Search Team” \cite{ref3,ref4,ref5}, observations of cosmic microwave background by COBE \cite{ref6} and WMAP \cite{ref7,ref8,ref9}, galaxy redshift surveys \cite{ref10}, etc., it is no matter of debate that the present universe is undergoing an accelerated expansion phase. Now, it is a challenge to the theoreticians to develop physically plausible gravity theories that match with these observations. Although the General Relativity (GR) theory has been incredibly successful in explaining the evolution of an expanding universe, the current stage of the universe's accelerating expansion has not been sufficiently explained by it. For the study of this scenario, cosmologists focused on modifying the already present gravity theories while simultaneously searching for new gravity theories. In the early attempts to achieve the accelerating expansion phase of the universe, the cosmological constant $\Lambda$, which was originally used by Einstein to keep the universe static and discarded after the discovery of the expanding universe, was reintroduced into the field equation derived from GR. The most accurate estimate that balances the cosmological constant's predictions in GR with observations is the $\Lambda$CDM standard cosmological model of the universe. A type of energy known as ``dark energy" that is uniformly distributed throughout space and has a constant energy density with an odd equation of state $p=-\rho$ is represented by the cosmological constant $\Lambda$ in the modified field equations.  The conventional cosmological model, known as the ``$\Lambda$CDM'' model \cite{ref11,ref12,ref13,ref14,ref15}, is the result of this dark energy, ordinary matter, and a hypothetical non-baryonic substance called Cold Dark Matter (CDM) that interacts solely gravitationally.\\
	
	Although the current accelerating expansion of the universe, large-scale structure formation, and many other observed universe attributes are explained by this standard model, however, as available cosmological data improve \cite{ref16,ref17,ref18,ref19,ref20}, a number of important problems with this conventional model arise that need to be resolved. So, the researchers were forced to look into potential modifications to existing relativity theories to address these issues. The generalized matter-geometry (GMG) coupling gravity theory inside the Riemannian framework is one of the numerous adaptations that we have examined.  According to theories of matter-geometry coupling, coupling is the process by which matter influences spacetime geometry and vice versa.  The metric tensor and the spacetime curvature represent the geometry in the Riemannian framework, whereas the energy-momentum tensor describes the matter field.  Additionally, studies of matter-geometric coupling theories using non-Riemannian geometric structures, such as Weyl geometry, have been conducted \cite{ref21}.  A gauge function, also called a displacement vector, is added to Lyra's geometry to alter the way matter interacts with spacetime geometry, giving rise to the concept of "matter-geometry coupling" \cite{ref22}.  When J. B. Jim\'{e}nez et al. \cite{ref23} applied the minimal coupling principle (MCP) to all standard model gauge fields and matter fields in a completely general (linear) affine geometry, they demonstrated the connection of matter and spacetime geometry.\\ 
	
	Buchdahl \cite{ref24} proposed a more comprehensive form of GR theory in 1970 by modifying the Einstein-Hilbert action $S=\int[\frac{R}{\kappa^{2}}+(L_{m})]\sqrt{-g}d^{4}x$ to $S=\int[\frac{f(R)}{\kappa^{2}}+(L_{m})]\sqrt{-g}d^{4}x$. Here the Lagrangian $L = R$ is replaced by $L = f(R)$ a general function of $R$. This $f(R)$ gravity is able to describe the behavior of massive test particles and the existence of a late-time cosmic acceleration without requiring dark energy or dark matter. This concept was utilized by researchers \cite{ref25, ref26, ref27, ref28, ref29, ref30} to resolve issues with the cosmological constant and to describe the current scenario of the universe's accelerated expansion. Early cosmological models that take into consideration both cosmic inflation and cosmic accelerations are provided by the modified $f(R)$ theory.\\
	
	By directly relating the matter Lagrangian $L_{m}$ to the Ricci scalar $R$, extended $f(R)$ gravity theories imply a connection between matter and curvature of space-time. An extra force orthogonal to the four-velocity is produced as a result of the coupling. Various forms of matter geometry coupling are utilized in literature. The action $S=\int[f_{1}(R)+(1+\lambda f_{2}(R))L_{m}]\sqrt{-g}d^{4}x$ was used by researchers in \cite{ref31,ref32,ref33} to determine the additional link between matter and curvature, which produced additional forces. A more general action for an $f(R)$-type modified gravity with an arbitrary coupling between matter and geometry was given by Harko \cite{ref34}: $S=\int[\frac{1}{2}f_{i}(R)+G(L_{m}) f_{2}(R)]\sqrt{-g}d^{4}x$, where $G(L_{m})$ is an arbitrary function of the matter Lagrangian density $L_m$, and $f_{i}(R), i = 1, 2$ are arbitrary functions of the Ricci scalar $R$. In their introduction of $f(R, L_{m})$ gravity theory, Harko and Lobo \cite{ref35} used the action $S=\int[f(R, L_{m})]\sqrt{-g}d^{4}x$ and the Lagrangian function $f$, an arbitrary function $R$ and $L_{m}$. The late-time cosmic acceleration scenario and physical phenomena of the known universe have been studied by several scientists using matter-geometry coupling theories in recent literature \cite{ref36,ref37,ref38,ref39,L25}. Using empirical constraints, we recently studied universe evolution in the $f(R,L_{m})$-gravity in \cite{ref40,ref41,ref42,ref43,ref44}. In $f(R,T)$ gravity theory, Harko et al. \cite{ref45} present an alternative coupling between matter and geometry. The action is given by $S=\int[f(R,T)+ L_{m}]\sqrt{-g}d^{4}x$, and $f(R,T)$ is an arbitrary function of $R$ and trace $T$ of the energy-momentum tensor (EMT) $T_{ij}$, ($T=g^{ij}T_{ij}$). In the literature, several investigations with different forms of ``$f(R,T)$'' function can be found \cite{ref46,ref47,ref48,ref49,ref50,ref51,ref52,ref53,ref54,ref55,ref56,ref57,ref58,ref59,ref60,ref61,ref62,ref63,ref64,ref65,ref66,ref67,ref68}.\\
	
	The $f(R,L_{m},T)$ gravity proposed by Haghani and Harko \cite{ref69} is a unified form of the $f(R)$, $f(R,L_{m})$, and $f(R,T)$ gravity theories. While $f(R)$, $f(R,L_{m})$, and $f(R,T)$ theories, among others, have their own advantages in describing the history of the universe, our present work concentrates on the more thorough and less studied $f(R,L_{m},T)$ gravity theory. In this unified gravity $f(R,L_{m},T)$, we examined the FLRW spacetime with a perfect fluid characterized by constant equation of state (EoS) parameters.   Spacetime is regarded as flat, homogeneous, and isotropic.   We recently examined constrained cosmological models \cite{ref70,ref71} within the framework of $f(R, L_{m},T)$ gravity, incorporating barotropic fluid and a non-linear $f(R, L_{m},T)$ function.   Pradhan et al. examined an accelerating universe model comprising dust within the framework of $f(R, T, L_{m})$ gravity in \cite{ref72}. In a more recent study, we have presented how the universe has evolved over cosmic time, assuming a simpler version of the $f(R, T, L_{m})$ function and constraining the model with the help of observational data \cite{ref73}.\\ 
	
	Dark energy notions can also be used to explain the current accelerated expansion of the universe at large, in addition to these modified matter-geometry coupling hypotheses. In this approach, the researchers theorized that in the universe, along with the usual matter, there is present a mysterious energy termed ` `dark energy," which exerts a very strong negative pressure and accelerates expansion, which is not possible by the normal cosmic matter. Although the strong energy condition $\rho+3 p\geq 0$ is violated by the considerable negative pressure, a closer examination of the data, however, shows that even the material that does not meet the weak energy condition $\rho + p \geq 0$, $\rho \ge{0}$, is acceptable with a high degree of confidence. Based on cosmological constraints, there are two types of dark energy models. With an EoS parameter $\omega_{de}> -1$, quintessence dark energy upholds the null energy requirement $\rho_{de}+p_{de}\geq 0$, whereas the phantom dark energy violates it with an EoS parameter $\omega_{de}< -1$ \cite{ref74,ref75,ref76}. Quintessence models $(\omega_{de}>-1)$ and phantom models $(\omega_{de}<-1)$ are separated by a third possibility, referred to as Quintom, in which the EoS parameter for dark energy $\omega_{de}$ may cross the barrier $\omega_{de}=-1$. Cai et al. \cite{ref77} developed an inflationary cosmology model with a big-bounce singularity instead of a big-bang singularity using the quintom scenario.\\
	
	Motivated by the above discussion of matter-geometry coupling theory, in this paper, we have investigated $f(R, T, L_{m})$ gravity considering a non-linear form of the $f(R, T, L_{m})$ function in $R$, $T$, and $L_{m}$ of the form $f(R, T, L_{m})=\alpha\,R+(\beta\,T+\gamma\,L_{m})R-\eta$ with $\alpha$, $\beta$, $\gamma$, and $\eta$ as arbitrary constants that have to be determined. We consider the ordinary matter energy density $\rho=\rho_{m}+\rho_{r}$ the sum of matter and radiation energy densities and solve the field equations to get the Hubble function $H(z)$. To estimate the best-fit values of model parameters, we have used MCMC analysis based on $31$ points of the Hubble function $H(z)$ and $1048$ points of apparent magnitudes $m(z)$ of the Pantheon sample of SNe Ia at $\sigma1$ and $\sigma2$ confidence levels. Though our present work is based on a matter-geometry coupling theory, it resembles the concepts of phantom dark energy.\\
	
	The paper is organized as follows: Section-1 contains introduction and some literature review while Section-2 presents the theory of $f(R,T,L_{m})$ gravity and field equations. In section-3, we have mention the solution of the field equations for function $f(R, T, L_{m})=\alpha\,R+(\beta\,T+\gamma\,L_{m})R-\eta$ and viability of the function $f(R, T, L_{m})$ while Section-4 contains the methodology of data analysis and estimation of cosmological parameters $H_{0}$, $\Omega_{m0}$, $\Omega_{r0}$, $\Omega_{\eta}$ and $\beta$. The cosmic 
	evolution history of the model is explored in section-5. In the section-6 we have performed some diagnostic tests on our model for its validity. Finally the conclusions are presented in section-7. 
		
\section{Modified field equations in $f(R,T,L_{m})$ theory}

To investigate the cosmic evolution of the universe in $f(R,T,L_{m})$ gravity theory, we take the action for $f(R,T,L_{m})$ theory of gravity of the form \cite{ref69}
\begin{equation}\label{eq1}
I=\frac{1}{16\pi}\int{[f(R,T,L_{m})+16\pi\,L_{m}]\sqrt{-g}d^{4}x},
\end{equation}
Here, $R$ represents the Ricci scalar, $T$ signifies the trace of $T_{ij}$, $L_{m}$ denotes the matter-Lagrangian density, and $f(R,T,L_{m})$ is a variable function of $R$, $T$, and $L_{m}$.  The principal physical rationale for considering this Lagrangian function is its non-triviality across all categories of matter fields, including radiation at $T=0$ \cite{ref69}.  The stress-energy momentum tensor of matter is defined as \cite{ref79}.
\begin{equation}\label{eq2}
  T_{ij}=-\frac{2}{\sqrt{-g}}\frac{\delta(\sqrt{-g}L_{m})}{\delta{g^{ij}}},
\end{equation}
and its trace given by $T=g^{ij}T_{ij}$, respectively.  Assuming that the Lagrangian density $L_{m}$ of matter is only a function of the metric tensor components $g_{ij}$, without dependence on their derivatives, we derive
\begin{equation}\label{eq3}
  T_{ij}=g_{ij}L_{m}-2\frac{\partial{L_{m}}}{\partial{g^{ij}}}.
\end{equation}
With regard to the metric tensor components $g^{ij}$, altering the gravitational field's action $I$ yields the following equation:
\begin{equation}\label{eq4}
  \delta{I}=\int\left[{f_{R}\delta{R}+f_{T}\frac{\delta{T}}{\delta{g^{ij}}}\delta{g^{ij}}+f_{L_{m}}\frac{\delta{L_{m}}}{\delta{g^{ij}}}\delta{g^{ij}}-\frac{1}{2}g_{ij}f\delta{g^{ij}}}-8\pi\,T_{ij}\delta{g^{ij}}\right]\sqrt{-g}d^{4}x,
\end{equation}
$f_{R}=\partial{f}/\partial{R}$, $f_{T}=\partial{f}/\partial{T}$, and $f_{L_{m}}=\partial{f}/\partial{L_{m}}$, respectively.  We derive, with respect to the Ricci scalar variation,
\begin{equation}\label{eq5}
  \delta{R}=\delta(g^{ij}R_{ij})=R_{ij}\delta{g^{ij}}+g^{ij}(\nabla_{l}\delta\Gamma^{l}_{ij}-\nabla_{j}\delta\Gamma^{l}_{il}),
\end{equation}
This is where $\nabla_{l}$ represents the covariant derivative with regard to the symmetric connection $\Gamma$ that is linked to the metric $g$.  The Christoffel symbols' variants produce
\begin{equation}\label{eq6}
  \delta{\Gamma^{l}_{ij}}=\frac{1}{2}g^{l\lambda p}(\nabla_{i}\delta{g_{jp}}+\nabla_{j}\delta{g_{p i}}-\nabla_{p}\delta{g_{ij}}),
\end{equation}
where the expression is given by the variation of the Ricci scalar, as
\begin{equation}\label{eq7}
  \delta{R}=R_{ij}\delta{g^{ij}}+g_{ij}\Box\delta{g^{ij}}-\nabla_{i}\nabla_{j}\delta{g^{ij}}.
\end{equation}
As a result of varying the action \eqref{eq1}, we get
\begin{equation}\label{eq8}
 \delta{R}=\int\left[f_{R}R_{ij}+(g_{ij}\Box-\nabla_{i}\nabla_{j})f_{R}+f_{T}\frac{\delta(g^{pq}T_{pq})}{\delta{g^{ij}}}+f_{L_{m}}\frac{\delta{L_{m}}}{\delta{g^{ij}}}-\frac{1}{2}g_{ij}f-8\pi\,T_{ij}\right]\delta{g^{ij}}\sqrt{-g}d^{4}x.
\end{equation}
where
\begin{equation}\label{eq9}
  \frac{\delta(g^{pq}T_{pq})}{\delta{g^{ij}}}=T_{ij}+\Theta_{ij},
\end{equation}
with
\begin{equation}\label{eq10}
  \Theta_{ij}=g^{pq}\frac{\delta T_{pq}}{\delta{g^{ij}}}=L_{m}g_{ij}-2T_{ij},
\end{equation}
for an ideal fluid matter supply.\\
The field equations of the $f(R,T,L_{m})$ gravity model are obtained by setting $\delta{I}=0$.
\begin{equation}\label{eq11}
  f_{R}R_{ij}-\frac{1}{2}[f-(f_{L_{m}}+2f_{T})L_{m}]g_{ij}+(g_{ij}\Box-\nabla_{i}\nabla_{j})f_{R}=8\pi\,T_{ij}+\frac{1}{2}(f_{L_{m}}+2f_{T})T_{ij}.
\end{equation}
On the contraction of Eq.~(\ref{eq11}), we obtain the following relationship:
\begin{equation}\label{eq12}
f_{R}R-2[f-(f_{L_{m}}+2f_{T})L_{m}]+3\Box f_{R}=8\pi\,T+\frac{1}{2}(f_{L_{m}}+2f_{T})T.
\end{equation}
The problem of ideal fluids, defined by an energy density $\rho$, pressure $p$, and four-velocity $u^{i}$, is more intricate due to the lack of a singular description for the matter Lagrangian.   In this paper, we assert that the stress-energy tensor of matter is denoted by
\begin{equation}\label{eq13}
  T_{ij}=(\rho+p)u_{i}u_{j}+pg_{ij},
\end{equation}
for the flat FLRW homogeneous and isotropic spacetime metric 
\begin{equation}\label{eq14}
	ds^{2}=-dt^{2}+a(t)^{2}(dx^{2}+dy^{2}+dz^{2}).
\end{equation}
$u_{i}u^{i}=-1$ and $u^{i}\nabla_{j}u_{i}=0$ are the constraints that the four-velocity $u_{i}$ must satisfy.\\
For a flat FLRW spacetime metric, the field equations can be written as
\begin{equation}\label{eq15}
	R_{0}^{0}-\frac{1}{2}R\delta_{0}^{0}=\frac{16\pi+f_{L_{m}}+2f_{T}}{2f_{R}}T_{0}^{0}+\frac{1}{2f_{R}}[f-Rf_{R}-(f_{L_{m}}+2f_{T})L_{m}-6H\dot{f}_{R}]\delta_{0}^{0},
\end{equation}
\begin{equation}\label{eq16}
	R_{1}^{1}-\frac{1}{2}R\delta_{1}^{1}=\frac{16\pi+f_{L_{m}}+2f_{T}}{2f_{R}}T_{1}^{1}+\frac{1}{2f_{R}}[f-Rf_{R}-(f_{L_{m}}+2f_{T})L_{m}+6H\dot{f}_{R}+2\ddot{f}_{R}]\delta_{1}^{1}.
\end{equation}
Here and onward the over dot denotes the ordinary derivative with respect to cosmic time.\\
\section{Model solutions}

We take into account the following type of the arbitrary function $f(R,T,L_{m})$ in order to study the cosmological features of the modified gravity that was proposed earlier:
\begin{equation}\label{eq17}
  f(R,T,L_{m})= \alpha\,R+\beta\,RT+\gamma\,RL_{m}-\eta,
\end{equation}
where $\alpha$, $\beta$, $\gamma$ and $\eta$ are arbitrary constants. Recently, we have studied this quadratic form of the Lagrangian function $f$ in \cite{ref70,ref71}. \\
The above form of $f$ gives
\begin{equation}\label{eq18}
	f_{R}=\alpha+\beta\,T+\gamma\,L_{m},~~~~ f_{T}=\beta\,R, ~~~~ f_{L_{m}}=\gamma\,R.
\end{equation}
Now from Eqs.\,\eqref{eq17} and \eqref{eq18}, we rewrite the Eqs.\,\eqref{eq15} and \eqref{eq16} as follows, respectively:
\begin{equation}\label{eq19}
	3H^{2}=8\pi\,G^{eff}\rho+\rho_{de},
\end{equation}
\begin{equation}\label{eq20}
	2\dot{H}+3H^{2}=-8\pi\,G^{eff}p-p_{de},
\end{equation}
where $G^{eff}$ denotes the effective gravitational constants, $\rho_{de}$ and $p_{de}$ are the dark energy density and pressure derived from matter-curvature coupling, respectively. These are defined as follows, respectively:
\begin{equation}\label{eq21}
	G^{eff}=\frac{16\pi+(2\beta+\gamma)R}{16\pi[\alpha+\beta\,T+\gamma\,L_{m}]},
\end{equation}
\begin{equation}\label{eq22}
	\rho_{de}=\frac{(2\beta+\gamma)R\,L_{m}+6H(\beta\dot{T}+\gamma\dot{L}_{m})+\eta}{2[\alpha+\beta\,T+\gamma\,L_{m}]},
\end{equation}
and
\begin{equation}\label{eq23}
	p_{de}=\frac{-(2\beta+\gamma)R\,L_{m}+6H(\beta\dot{T}+\gamma\dot{L}_{m})+2(\beta\ddot{T}+\gamma\ddot{L}_{m})-\eta}{2[\alpha+\beta\,T+\gamma\,L_{m}]}.
\end{equation}
The equation of motion is obtained as
\begin{equation}\label{eq24}
	\dot{\rho}+3H(\rho+p)+\frac{\dot{G}^{eff}}{G^{eff}}\rho=-\frac{1}{8\pi\,G^{eff}}[\dot{\rho}_{de}+3H(\rho_{de}+p_{de})].
\end{equation}
The equation of state parameter for dark sector is obtained from Eqs.\,\eqref{eq22} and \eqref{eq23} as below
\begin{equation}\label{eq25}
	\omega_{de}=-1+\frac{12H(\beta\dot{T}+\gamma\dot{L}_{m})+2(\beta\ddot{T}+\gamma\ddot{L}_{m})}{(2\beta+\gamma)R\,L_{m}+6H(\beta\dot{T}+\gamma\dot{L}_{m})+\eta}.
\end{equation}

Now, to get an exact solution of the field equations \eqref{eq19} and \eqref{eq20}, we use $T=-\rho+3p$, $L_{m}=-\rho$ and $R=6(\dot{H}+2H^{2})$ in Eq.\,\eqref{eq19}, we have
\begin{equation}\label{eq26}
	6H^{2}=\frac{16\pi\rho+\eta}{\alpha-(\beta+\gamma)[\rho+(1+z)\rho']+3\beta[p+(1+z)p']},
\end{equation}
where prime denotes the ordinary derivatives with respect to redshift $z$.\\
Although this modified theory is not conservative, it does not meet the general energy conservation equation.  However, we assume the perfect fluid source as matter and radiations that may be derived from $\dot{\rho}_{m}+3H\rho_{m}=0$ and $\dot{\rho}_{r}+4H\rho_{r}=0$,
\begin{equation}\label{eq27}
	\rho=\rho_{m}+\rho_{r},\,\,\,\,p=\omega_{m}\rho_{m}+\omega_{r}\rho_{r},
\end{equation}
where $\rho_{m}$ and $\rho_{r}$ denote the energy densities corresponding to matter and radiation, respectively, and $\omega_{m}$ and $\omega_{r}$ are corresponding equation of state parameters. The total energy density of the perfect fluid is obtained as $\rho=\rho_{m0}a^{-3}+\rho_{r0}a^{-4}$ and pressure $p=\frac{1}{3}\rho_{r0}a^{-4}$ with standard convention $a_{0}=1$ the current value of scale factor $a(t)$, where the EoS parameter value for matter $\omega_{m}=0$ and for radiation $\omega_{r}=\frac{1}{3}$ while $\rho_{m0}$ and $\rho_{r0}$ are the current values of corresponding energy densities, respectively. Here, we have used the relationship between scale factor $a(t)$ and redshift $z$, $a_{0}/a(t)=1+z$, as given in \cite{ref89}. Thus, the energy density and pressure for matter source fluid are derived as below, respectively:
\begin{equation}\label{eq28}
	\rho(z)=\rho_{m0}(1+z)^{3}+\rho_{r0}(1+z)^{4},\,\,\,\,p(z)=\frac{1}{3}\rho_{r0}(1+z)^{4}.
\end{equation}
Now, using Eq.\,\eqref{eq28} in \eqref{eq26}, we obtain the Hubble function as below
\begin{equation}\label{eq29}
	H(z)=\sqrt{\frac{\Omega_{m0}(1+z)^{3}+\Omega_{r0}(1+z)^{4}+\Omega_{\eta}}{\frac{\alpha}{H_{0}^{2}}-\frac{3(\beta+\gamma)}{2\pi}\Omega_{m0}(1+z)^{3}-\frac{15\gamma}{8\pi}\Omega_{r0}(1+z)^{4}}},
\end{equation}
where $\Omega_{m0}=\frac{8\pi\rho_{m0}}{3H_{0}^{2}}$, $\Omega_{r0}=\frac{8\pi\rho_{r0}}{3H_{0}^{2}}$ and $\Omega_{\eta}=\frac{\eta}{6H_{0}^{2}}$. For the validity of this equation \eqref{eq29}, we substitute $z=0$ which gives the following relationship:
\begin{equation}\label{eq30}
	\Omega_{m0}+\Omega_{r0}+\Omega_{\eta}=\alpha-\frac{3(\beta+\gamma)H_{0}^{2}}{2\pi}\Omega_{m0}-\frac{15\gamma\,H_{0}^{2}}{8\pi}\Omega_{r0}.
\end{equation}
By removing the arbitrary constant $\alpha$ from Eqs.\,\eqref{eq29} and \eqref{eq30}, we derive the Hubble function as follows
\begin{equation}\label{eq31}
	H(z)=H_{0}\sqrt{\frac{\Omega_{m0}(1+z)^{3}+\Omega_{r0}(1+z)^{4}+\Omega_{\eta}}{\Omega_{m0}+\Omega_{r0}+\Omega_{\eta}+\frac{3(\beta+\gamma)H_{0}^{2}}{2\pi}\Omega_{m0}[1-(1+z)^{3}]+\frac{15\gamma\,H_{0}^{2}}{8\pi}\Omega_{r0}[1-(1+z)^{4}]}},
\end{equation}
where $H_{0}$, $\Omega_{m0}$, $\Omega_{r0}$, $\Omega_{\eta}$, $\beta$, and $\gamma$ are parameters to be constrained with the help of observational datasets for Hubble function.\\

\subsection{Viability of the $f(R, T, L_{m})$ function}

Here, we discuss the viability of the considered non-linear form of the function $f(R, T, L_{m})=\alpha\,R+(\beta\,T+\gamma\,L_{m})R-\eta$ which by using $T=-\rho+3p$, $L_{m}=-\rho$, can be expressed as
\begin{equation}\label{eq32}
	f(R, T, L_{m})=\alpha\,R-(\beta+\gamma)R\rho+3\beta\,R\,p-\eta.
\end{equation}
The partial derivative of the above function with respect to Ricci scalar $R$ and energy density $\rho$ are obtained, respectively, as below:
\begin{equation}\label{eq33}
	\frac{\partial f}{\partial R}=\alpha-(\beta+\gamma)\rho+3\beta\,p,
\end{equation}
\begin{equation}\label{eq34}
	\frac{\partial f}{\partial \rho}=-(\beta+\gamma)R+3\beta\,R\,\frac{\partial p}{\partial \rho}.
\end{equation}
We have considered ordinary matter fluid for which $\rho\ge0$, $p\ge0$, and we can see that the Ricci scalar $R>0$. Furthermore, it is clear that $\frac{\partial p}{\partial \rho}\ge0$. For the viability of the model, the following inequalities should be satisfied by the $f(R, T, L_{m})$ function: $f>0$, $\frac{\partial f}{\partial R}>0$, and $\frac{\partial f}{\partial \rho}>0$. One can observe that these inequalities will be satisfied for $\alpha>0$, $\beta<0$, $\gamma<0$, and $\eta>0$. In the next section, we will use these conditions on arbitrary constants $\alpha$, $\beta$, $\gamma$, and $\eta$ in the estimation of observational constraints on model parameters.

\section{Estimation of cosmological parameters $H_{0}$, $\Omega_{m0}$, $\Omega_{r0}$, $\Omega_{\eta}$ and $\beta$}

In this section, we have estimated the values of cosmological parameters $H_{0}$, $\Omega_{m0}$, $\Omega_{r0}$, $\Omega_{\eta}$, and $\beta$ for which the above-derived model is valid, viable, and acceptable. We constrained these parameters using Eq.\,\eqref{eq31}, which presents the Hubble function $H(z)$ in relation to these parameters. A ``Monte Carlo Markov Chain (MCMC)" joint analysis was conducted on $31$ ``cosmic chronometer (CC) Hubble data", derived from the differential age method, alongside $1048$ apparent magnitudes $m(z)$ from the ``Pantheon sample of SNe Ia". For MCMC analysis, we use the emcee software available freely at \cite{ref90}.

\subsection{Hubble data}

The data points of the Hubble constant are useful to understand the expansion rate of the universe. The Hubble parameter links the observational and theoretical models, and so, we have derived the Hubble function from the field equations. We will estimate the parameters $H_{0}$, $\Omega_{m0}$, $\Omega_{r0}$, $\Omega_{\eta}$, and $\beta$ involved in $H(z)$ using $31$ CC datasets \cite{ref91,ref92}, in the MCMC analysis applying the $\chi^{2}$ formula as given below:
\begin{equation}\label{eq35}
	\chi_{CC}^{2}=\sum_{i=1}^{i=31}\frac{[H_{ob}(z_{i})-H_{th}(\theta, z_{i})]^{2}}{\sigma_{H(z_{i})}^{2}},
\end{equation}
where $\theta$ represents the set of unknown parameters $H_{0}$, $\Omega_{m0}$, $\Omega_{r0}$, $\Omega_{\eta}$ and $\beta$ which we have to estimate, $H_{ob}$ denotes the observed values of $H(z)$ at $z=z_{i}$ and $H_{th}$ is the theoretical value of $H(z)$ at $z=z_{i}$ while $\sigma_{H(z_{i})}$ denotes the standard deviations in Hubble data points.\\
\begin{figure}[H]
	\centering
	\includegraphics[width=10cm,height=10cm,angle=0]{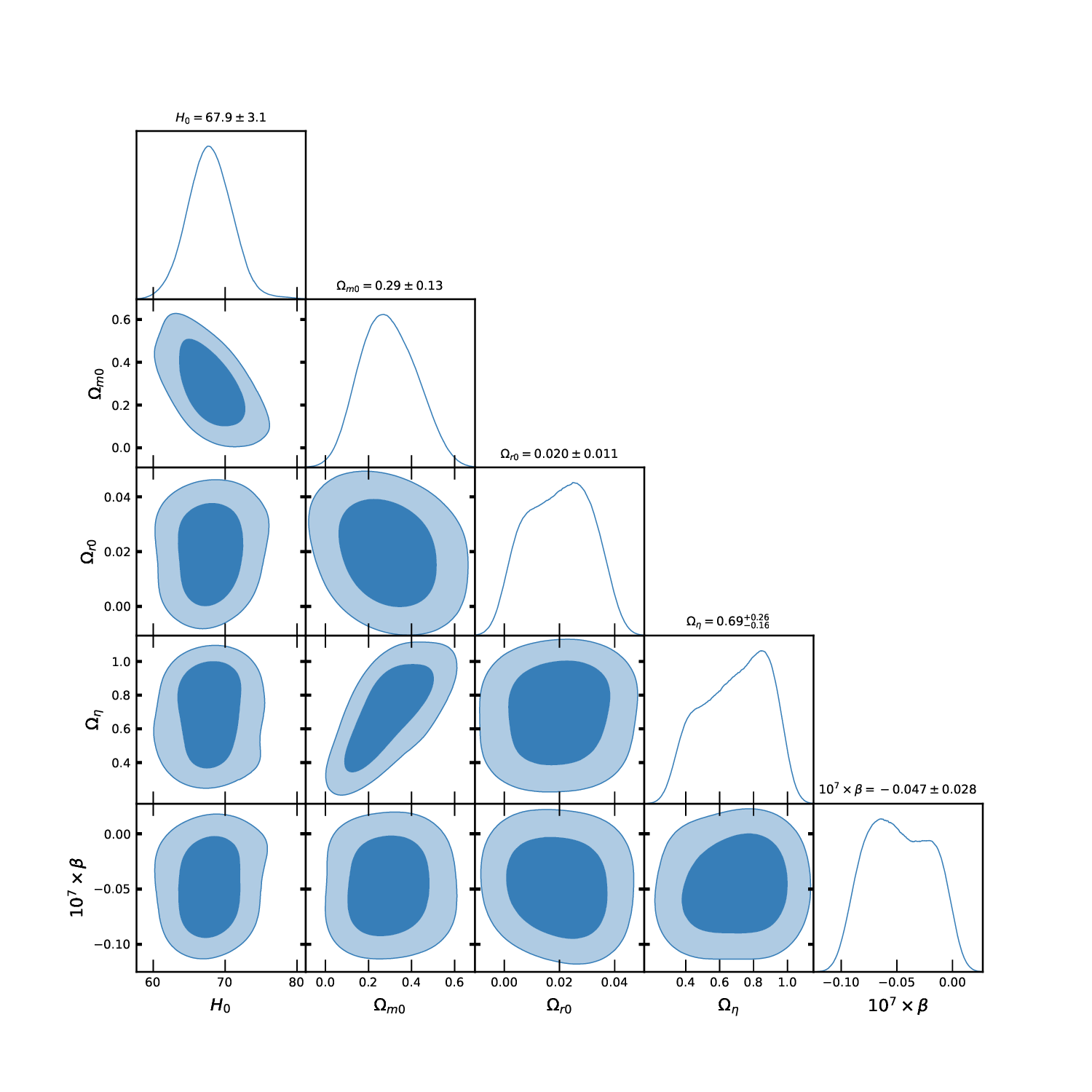}
	\caption{The contour plots of $H_{0}$ $\Omega_{m0}$, $\Omega_{r0}$, $\Omega_{\eta}$ and $\beta$ at $1-\sigma, 2-\sigma$ confidence level in MCMC analysis of CC datasets.}
\end{figure}
Figure 1 depicts $\sigma1$ and $\sigma2$ confidence level contour plots of $H_{0}$, $\Omega_{m0}$, $\Omega_{r0}$, $\Omega_{\eta}$, and $\beta$ using the CC datasets. In the MCMC analysis of $31$ CC datasets, we use a wide range of priors with suitable feasible initial values for each parameter. The obtained values of parameters in MCMC analysis are shown in Figure 1 and Table 1. We measured the value of the Hubble constant $H_{0}=67.9\pm3.1$ km/s/Mpc with matter and radiation energy density parameters $\Omega_{m0}=0.29\pm0.13$ and $\Omega_{r0}=0.020\pm0.011$, respectively. We measured $\Omega_{\eta}=0.69_{-0.16}^{+0.26}$ and the coupling constant $\beta=(-0.047\pm0.028)\times10^{-7}$. One can see that the parameters $\beta$ and $\gamma$ show degeneracy. Hence, in our analysis, we assume $\beta=\gamma$ since both coupling constants are used in coupling matter and curvature multiplied terms. We have measured the value of $\alpha$ for CC datasets as $\alpha=1.03$, which is acceptable.

\subsection{Pantheon data}

The universe's expansion history can be characterized using SNe Ia data. In this study, we utilize $1048$ data points of apparent magnitude $m(z)$ from the Pantheon sample, covering the redshift range $0.01 \le z \le 2.26$, as referenced in \cite{ref93}, to impose observational constraints on the model parameters. We derive the $m(z)$ \cite{ref93,ref94}, as
\begin{equation}\label{eq36}
	m(z)=M+ 5~\log_{10}\left(\frac{D_{L}}{Mpc}\right)+25,
\end{equation}
where $M$ represents the absolute apparent magnitude, and the luminosity distance $D_{L}$ is defined as
\begin{equation}\label{eq37}
	D_{L}=c(1+z)\int^z_0\frac{dz'}{H(z')}.
\end{equation}
Let us define \( h(z) = \frac{H(z)}{H_{0}} \). Consequently, the preceding equation \eqref{eq37} can be expressed as
\begin{equation}\label{eq38}
	D_{L}=\frac{c}{H_{0}}(1+z)\int^z_0\frac{dz'}{h(z')}.
\end{equation}
We define $D_{L}=\frac{c}{H_{0}}d_{L}$, where $d_{L}$ is a dimensionless quantity. Consequently, the apparent magnitude $m(z)$ in Eq.\eqref{eq36} can be reformulated as
\begin{equation}\label{eq39}
	m(z)=\mathcal{M}+5\log_{10}{d_{L}},
\end{equation}
We establish a dimensionless parameter $\mathcal{M}=25+M+5\log_{10}\left(\frac{c/H_{0}}{Mpc}\right)$ by integrating two degenerate parameters, $H_{0}$ and $M$, which remain constant within the $\Lambda$CDM framework \cite{ref93,ref94}.  The expression can be formulated as $\mathcal{M}=M-5\log_{10}(h)+42.39$, where $H_{0}=h\times100\,km/s/Mpc$.  The analysis of the Pantheon data is conducted using the following $\chi^{2}$ formula, as referenced in \cite{ref93,ref95,ref96,ref97}:
\begin{equation}\label{eq40}
	\chi^{2}_{P}=V_{P}^{i}C_{ij}^{-1}V_{P}^{j},
\end{equation}
where the difference between the observed $m_{ob}(z_{i})$ and the theoretical $m_{th}(\theta, z_{i})$, as stated in equation \eqref{eq39}, is represented by the expression $V_{P}^{i}$.\\

The following $\chi^{2}$ formula is used for the joint analysis of the $31$ Hubble function CC datasets and the $1048$ Pantheon datasets of apparent magnitude.
\begin{equation}\nonumber
	\chi^{2}_{CC+P}=\chi^{2}_{CC}+\chi^{2}_{P}.
\end{equation}

\begin{table}[H]
	\centering
	\begin{tabular}{|c|c|c|c|}
		\hline
		
		Parameter          & Prior         & CC                        & CC$+$Pantheon  \\
		\hline
		$\Omega_{m0}$      & $(0, 0.6)$    & $0.29\pm0.13$             & $0.266_{-0.091}^{+0.11}$  \\
		$\Omega_{r0}$      & $(0, 0.04)$   & $0.020\pm0.011$           & $0.020_{-0.012}^{+0.013}$       \\
		$\Omega_{\eta}$    & $(0, 1)$      & $0.69_{-0.16}^{+0.26}$    & $0.70_{-0.14}^{+0.27}$    \\
		$10^{7}\times\beta$& $(-0.1, 0)$   & $-0.047\pm0.028$          & $-0.052_{-0.039}^{+0.025}$      \\
		$H_{0}$            & $(50, 100)$   & $67.9\pm3.1$              & $68.6\pm1.9$         \\
		$\mathcal{M}$      & $(23, 24)$    & -                         & $23.809\pm0.011$  \\
		$\chi^{2}$         &  -            & $14.4936$                 & $1041.0699$             \\
		\hline
	\end{tabular}
	\caption{The MCMC estimates.}\label{T1}
\end{table}
%
For the CC+Pantheon datasets, Figure 2 illustrates the contour plots of $H_{0}$, $\Omega_{m0}$, $\Omega_{r0}$, $\Omega_{\eta}$, and $\beta$ at the $\sigma1$ and $\sigma2$ confidence levels.  We employ a diverse selection of priors with plausible initial values for each parameter in the MCMC joint analysis of the $31$ CC and $1048$ pantheon datasets.  Figure 2 and Table 1 illustrate the estimated outcomes of the MCMC analysis.  Our measurements of the Hubble constant $H_{0}=68.6\pm1.9$ km/s/Mpc were conducted using the matter and radiation energy density parameters $\Omega_{m0}=0.266_{-0.091}^{+0.11}$ and $\Omega_{r0}=0.020_{-0.012}^{+0.013}$, respectively. We measured $\Omega_{\eta}=0.70_{-0.14}^{+0.27}$ and coupling constant $\beta=(-0.052_{-0.039}^{+0.025})\times10^{-7}$. One can see that the parameters $\beta$ and $\gamma$ show degeneracy. Hence, in our analysis, we assume $\beta=\gamma$ since both coupling constants are used in coupling matter and curvature multiplied terms. We have measured the value of $\alpha$ for CC+Pantheon datasets as $\alpha=0.986$, which is acceptable.

\begin{figure}[H]
	\centering
	\includegraphics[width=10cm,height=10cm,angle=0]{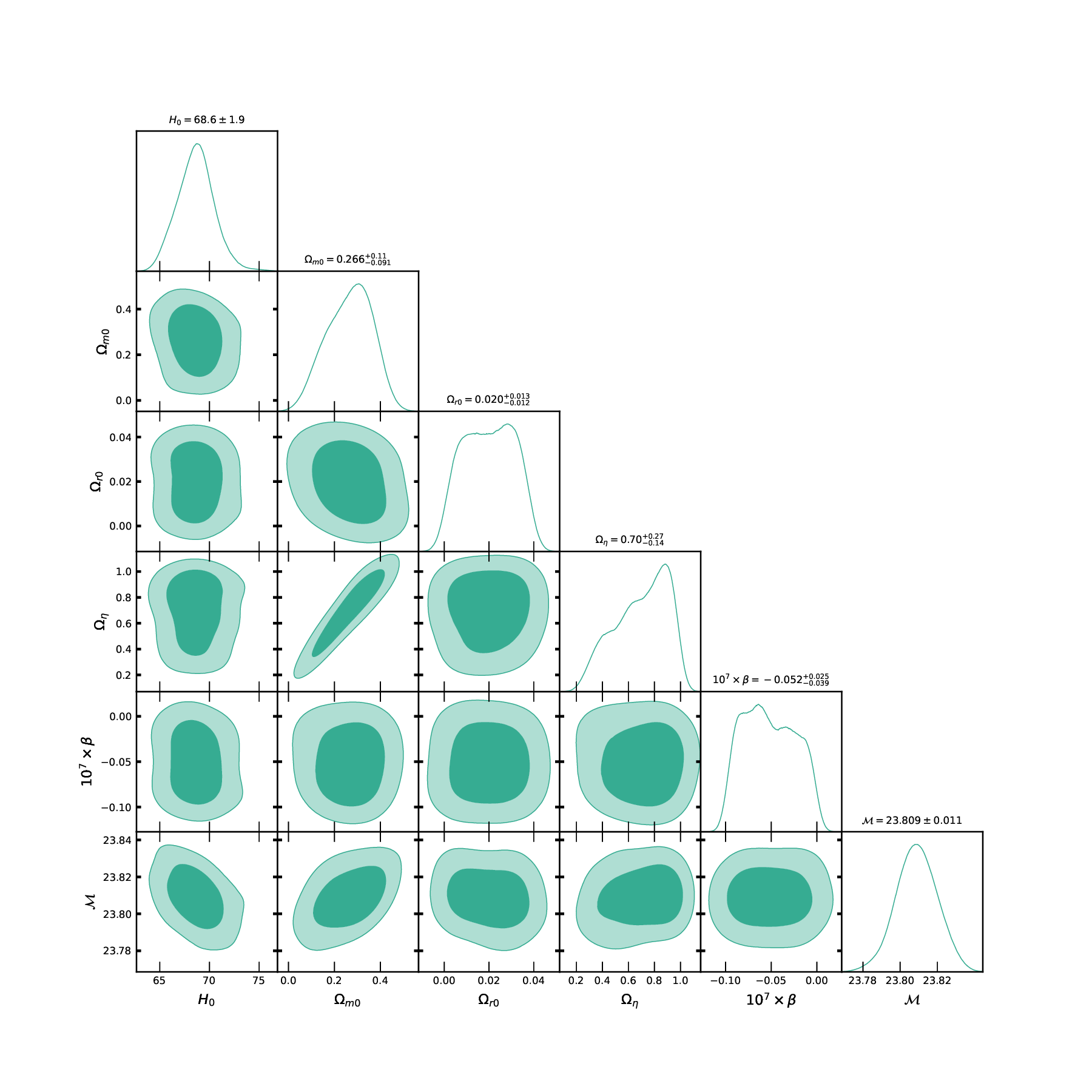}
	\caption{The contour plots of $H_{0}$ $\Omega_{m0}$, $\Omega_{r0}$, $\Omega_{\eta}$, $\beta$ and $\mathcal{M}$ at $1-\sigma, 2-\sigma$ confidence level in MCMC analysis of CC+Pantheon datasets.}
\end{figure}

\section{Cosmic evolution history of the universe}

To explore the expansion history of the model, in this section, we have discussed some cosmological parameters derived from the field equations of the generalized matter-curvature gravity theory using the assumed non-linear Lagrangian function $f(R, T, L_{m})=\alpha\,R+\beta\,RT+\gamma\,RL_{m}-\eta$. We analyzed the behavior of the cosmological parameters using the estimated values from Table \ref{T1} in the previous section.\\

First, we have derived the deceleration parameter $q=-1+\frac{d}{dt}\left(\frac{1}{H}\right)$ (using the Hubble function \eqref{eq31}), which is given as:
\begin{multline}\label{eq41}
	q(z)=-1+\frac{1}{2}\left[\frac{3\Omega_{m0}(1+z)^{3}+4\Omega_{r0}(1+z)^{4}}{\Omega_{m0}(1+z)^{3}+\Omega_{r0}(1+z)^{4}+\Omega_{\eta}}\right.\\\left.+\frac{\frac{9(\beta+\gamma)H_{0}^{2}}{2\pi}\Omega_{m0}(1+z)^{3}+\frac{15\gamma\,H_{0}^{2}}{2\pi}\Omega_{r0}(1+z)^{4}}{\Omega_{m0}+\Omega_{r0}+\Omega_{\eta}+\frac{3(\beta+\gamma)H_{0}^{2}}{2\pi}\Omega_{m0}[1-(1+z)^{3}]+\frac{15\gamma\,H_{0}^{2}}{8\pi}\Omega_{r0}[1-(1+z)^{4}]}\right].
\end{multline}
Figure 3 depicts the variation of $q(z)$ over redshift $z$ (as expressed by Eq.\,\eqref{eq41}). From the figure, we can observe that $q(z)$ is an increasing function of $z$. It is observed that $q\to-1$ as $z\to-1$, indicating that the cosmos will experience accelerated expansion in the far future. Also, it is shown that $q\to0.6$ as $z\to3$, demonstrating that the early universe was experiencing a decelerated expansion. The transition from decelerated to accelerated phase (ie. from $q > 0$ to $q <0$) took place at $z_{t}=0.5965$ for CC datasets and $z_{t}=0.6177$ for the joint datasets of $CC+Pantheon$.  We have also estimated the present value of $q(z)$ for CC datasets as $q_{0}=-0.5388$ and along CC+Pantheon data, $q_{0}=-0.5547$, which are negative that reveals that the present universe is undergoing an accelerating phase. These estimated values of $q_{0}$ and $z_{t}$ are compatible with the measurement from observed universe data in several studies \cite{ref98,ref99,ref100,ref101,ref102,ref103,ref104,ref105,ref106,ref107}.\\

\begin{figure}[H]
	\centering
	\includegraphics[width=10cm,height=8cm,angle=0]{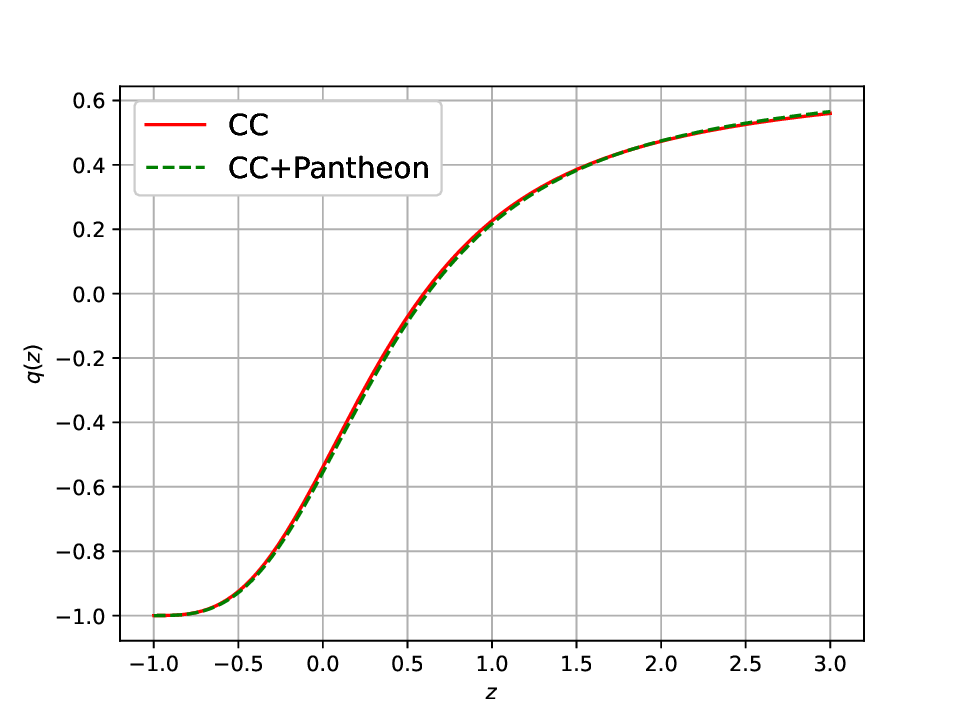}
	\caption{The variation of deceleration parameter $q(z)$ versus $z$.}
\end{figure}

Now, the effective EoS parameter $\omega_{eff}$ for the derived model can be defined in terms of the deceleration parameter $q$ as given below.
\begin{equation}\label{eq42}
	\omega_{eff}=\frac{2q-1}{3},
\end{equation}
and using Eq.\,\eqref{eq41} in \eqref{eq42}, we have
\begin{multline}\label{eq43}
	\omega_{eff}=-1+\frac{1}{3}\left[\frac{3\Omega_{m0}(1+z)^{3}+4\Omega_{r0}(1+z)^{4}}{\Omega_{m0}(1+z)^{3}+\Omega_{r0}(1+z)^{4}+\Omega_{\eta}}\right.\\\left.+\frac{\frac{9(\beta+\gamma)H_{0}^{2}}{2\pi}\Omega_{m0}(1+z)^{3}+\frac{15\gamma\,H_{0}^{2}}{2\pi}\Omega_{r0}(1+z)^{4}}{\Omega_{m0}+\Omega_{r0}+\Omega_{\eta}+\frac{3(\beta+\gamma)H_{0}^{2}}{2\pi}\Omega_{m0}[1-(1+z)^{3}]+\frac{15\gamma\,H_{0}^{2}}{8\pi}\Omega_{r0}[1-(1+z)^{4}]}\right].
\end{multline}
Equation \eqref{eq43} expresses the effective EoS parameter and its behavior with respect to redshift.  $z$ is depicted in Figure 4.  The graphic shows that $\omega_{eff}$ is an increasing function of $z$, with $\omega_{eff}\to-1$ as $z\to-1$ and $\omega_{eff}\to0.05$ as $z\to3$, indicating that the cosmos was dominated by matter in the beginning and dark energy (formed from matter-curvature) in the end.  We measure the present value of $\omega_{eff}=-0.6925$ for CC datasets and $\omega_{eff}=-0.7031$ for joint datasets CC+Pantheon, which are consistent with observations.  The model's effective energy density and pressure can be defined as follows: 
\begin{equation}\label{eq44}
	\rho_{eff}=\frac{3H^{2}}{8\pi\,G^{eff}},\,\,\,\,p_{eff}=\frac{2(1+z)HH'-3H^{2}}{8\pi\,G^{eff}},
\end{equation}
where $G^{eff}$ is defined as effective gravitational constant and derived as
\begin{equation}\label{eq45}
	G^{eff}=\frac{16\pi-6(2\beta+\gamma)(1+z)HH'+12(2\beta+\gamma)H^{2}}{16\pi\alpha-6H_{0}^{2}[(\beta+\gamma)\Omega_{m0}(1+z)^{3}+\gamma\Omega_{r0}(1+z)^{4}]}.
\end{equation}
\begin{figure}[H]
	\centering
	\includegraphics[width=10cm,height=8cm,angle=0]{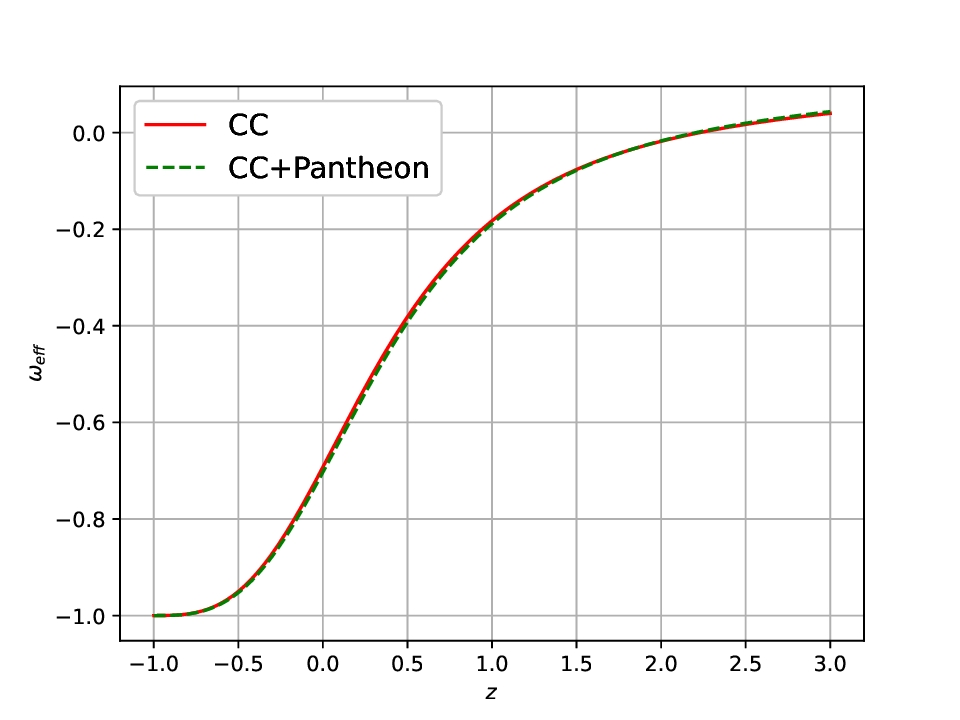}
	\caption{The variation of effective EoS parameter $\omega_{eff}$ versus $z$.}
\end{figure}
The dark energy density $\rho_{de}$ and pressure $p_{de}$ are defined in Eqs.\,\eqref{eq22} and \eqref{eq23}, respectively. We can rewrite these Eqs.\,\eqref{eq22} and \eqref{eq23}, respectively, as
\begin{equation}\label{eq46}
	\rho_{de}=\frac{6(2\beta+\gamma)HH'[\rho_{m0}(1+z)^{4}+\rho_{r0}(1+z)^{5}]-12(\beta-\gamma)\rho_{m0}H^{2}(1+z)^{3}-12(2\beta-\gamma)\rho_{r0}H^{2}(1+z)^{4}+\eta}{2[\alpha-(\beta+\gamma)\rho_{m0}(1+z)^{3}-\gamma\rho_{r0}(1+z)^{4}]},
\end{equation}
\begin{equation}\label{eq47}
	p_{de}=\frac{-2[3(5\beta+3\gamma)\rho_{m0}(1+z)^{4}+2(6\beta+5\gamma)\rho_{r0}(1+z)^{5}]HH'+8[3(2\beta+\gamma)\rho_{m0}(1+z)^{3}+2(3\beta+\gamma)\rho_{r0}(1+z)^{4}]H^{2}-\eta}{2[\alpha-(\beta+\gamma)\rho_{m0}(1+z)^{3}-\gamma\rho_{r0}(1+z)^{4}]},
\end{equation}
where prime denotes the ordinary derivatives with respect to $z$. The Hubble function $H(z)$ is given by Eq.\,\eqref{eq31} and its first derivative is obtained as
\begin{multline}\label{eq48}
	\frac{H'}{H}=\frac{1}{2}\left[\frac{3\Omega_{m0}(1+z)^{2}+4\Omega_{r0}(1+z)^{3}}{\Omega_{m0}(1+z)^{3}+\Omega_{r0}(1+z)^{4}+\Omega_{\eta}}\right.\\\left.+\frac{[9(\beta+\gamma)\Omega_{m0}(1+z)^{2}+15\gamma\Omega_{r0}(1+z)^{3}]H_{0}^{2}}{2\pi[\Omega_{m0}+\Omega_{r0}+\Omega_{\eta}+\frac{3(\beta+\gamma)H_{0}^{2}}{2\pi}\Omega_{m0}[1-(1+z)^{3}]+\frac{15\gamma\,H_{0}^{2}}{8\pi}\Omega_{r0}[1-(1+z)^{4}]]}\right].
\end{multline}
From Eqs.\,\eqref{eq46} and \eqref{eq47}, we derive the dark energy EoS parameter $\omega_{de}$ as below:
\begin{equation}\label{eq49}
	\omega_{de}=-1+\frac{3(\beta+\gamma)\Omega_{m0}(1+z)^{3}+\frac{8}{3}\gamma\Omega_{r0}(1+z)^{4}-(\beta+\gamma)\Omega_{m0}(1+z)^{4}\frac{H'}{H}-\frac{4}{3}\gamma\Omega_{r0}(1+z)^{5}\frac{H'}{H}}{(2\beta+\gamma)\frac{H'}{H}[\Omega_{m0}(1+z)^{4}+\Omega_{r0}(1+z)^{5}]-(\beta-\gamma)\Omega_{m0}(1+z)^{3}+4(\beta+\gamma)\Omega_{r0}(1+z)^{4}+\frac{8\pi\Omega_{\eta}}{3H^{2}}}.
\end{equation}
Figure 5 shows the evolution of the dark energy EoS parameter as given in Eq.\,\eqref{eq49}.  Figure 5 shows that $\omega_{de}$ is a decreasing function of $z$ across the interval $[-1, 3]$, with $\omega_{de}\to-1$ as $z\to-1$, indicating the model's late-time inclination towards the $\Lambda$CDM model.  The current value of dark energy EoS parameter is $\omega_{de}=-1.0000061047$ for CC data and $\omega_{de}=-1.000005844$ for the joint datasets of CC+Pantheon.\\
\begin{figure}[H]
	\centering
	\includegraphics[width=10cm,height=8cm,angle=0]{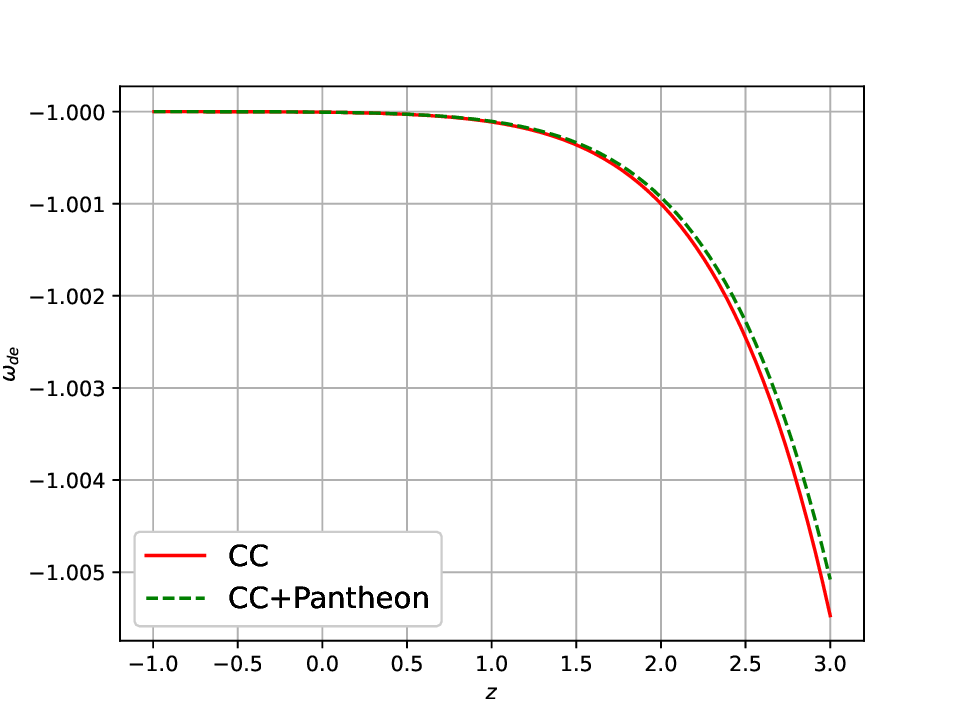}
	\caption{The variation of dark energy EoS parameter $\omega_{de}$ versus $z$.}
\end{figure}

\begin{figure}[H]
	\centering
	a.\includegraphics[width=8cm,height=7cm,angle=0]{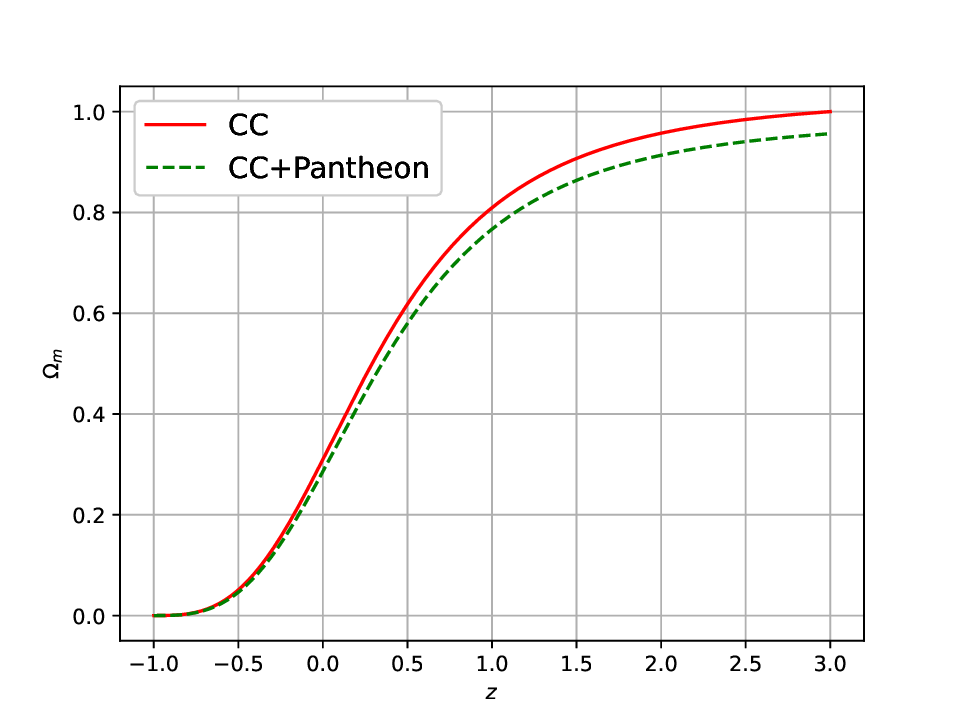}
	b.\includegraphics[width=8cm,height=7cm,angle=0]{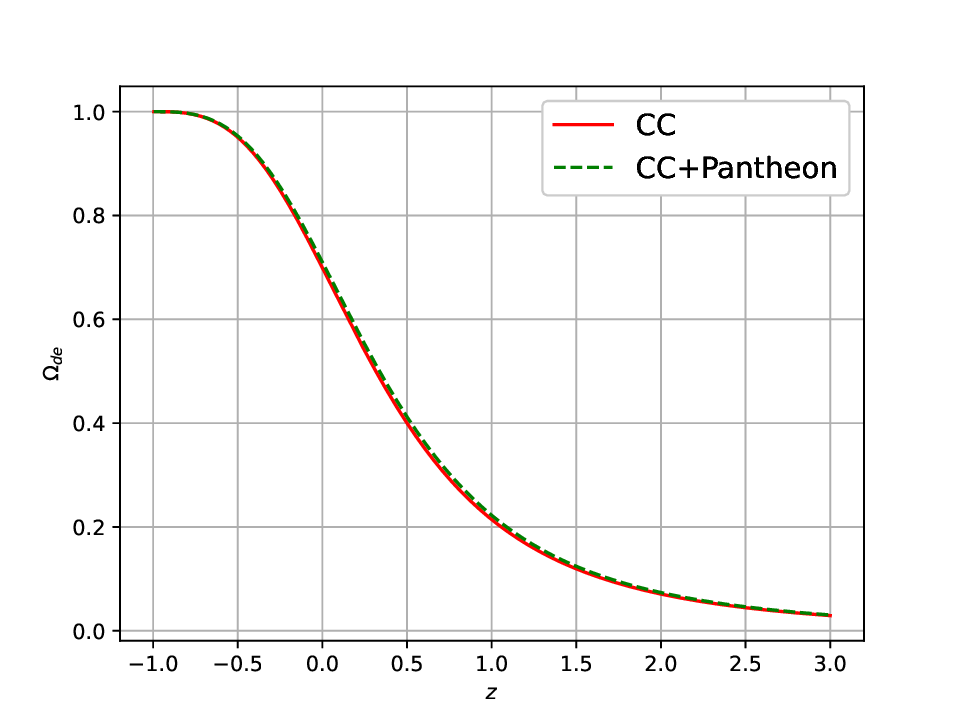}
	\caption{The variation of energy density parameters $\Omega_{m}$ and $\Omega_{de}$ versus $z$.}
\end{figure}
The matter energy density parameter $\Omega_{m}$ and dark energy density parameter $\Omega_{de}$ can be defined as
\begin{equation}\label{eq50}
	\Omega_{m}=\frac{8\pi\,G^{eff}\rho}{3H^{2}},\,\,\,\,\Omega_{de}=\frac{\rho_{de}}{3H^{2}}.
\end{equation}
The graphical presentation of $\Omega_{m}$ and $\Omega_{de}$ are Figures 6a and 6b, respectively. Figures 6a and 6b show that the dark energy density parameter $\Omega_{de}$ decreases with $z$, while $\Omega_{m}$ increases.  Figure 6a shows that the matter energy density parameter $\Omega_{m}\to0$ at late-time universe and $\Omega_{m}\to1$ as $z\to3$ in early universe evolution, while Figure 6b shows that the dark energy density parameter $\Omega_{de}\to1$ at late-time universe and $\Omega_{de}\to0$ as $z\to\infty$ in early time.  The current value of the matter energy density parameter is $\Omega_{m}\approx0.31$ for CC data and $\Omega_{m}\approx0.29$ for joint datasets. The current value of the dark energy density parameter is $\Omega_{de}\approx0.70$ for CC datasets and $\Omega_{de}\approx0.71$ for joint analysis.  These results are consistent with recent observations.\\

Next, we estimated the age of the universe using the following formula:
\begin{equation}\label{eq51}
	t_{0}-t=\int_{0}^{z}\frac{dz'}{(1+z')H(z')}.
\end{equation}
We denote $\frac{H(z')}{H_{0}}=h(z')$ in Eq.\,\eqref{eq51}, we can rewrite \eqref{eq51} as
\begin{equation}\label{eq52}
	(t_{0}-t)H_{0}=\int_{0}^{z}\frac{dz'}{(1+z')h(z')}.
\end{equation}
Using Eq.\,\eqref{eq31} in \eqref{eq52}, we have
\begin{equation}\label{eq53}
	t_{0}H_{0}=\lim_{t\to0}(t_{0}-t)H_{0}=\lim_{z\to\infty}\int_{0}^{z}\frac{dz'}{(1+z')\sqrt{\frac{\Omega_{m0}(1+z)^{3}+\Omega_{r0}(1+z)^{4}+\Omega_{\eta}}{\Omega_{m0}+\Omega_{r0}+\Omega_{\eta}+\frac{3(\beta+\gamma)H_{0}^{2}}{2\pi}\Omega_{m0}[1-(1+z)^{3}]+\frac{15\gamma\,H_{0}^{2}}{8\pi}\Omega_{r0}[1-(1+z)^{4}]}}}.
\end{equation}
Equation \eqref{eq53} states that as $t\to0$ and $z\to\infty$, $(t_{0}-t)H_{0}\to t_{0}H_{0}$, resulting in the universe's current age.  For CC datasets, we estimate $t_{0}H_{0}=0.9630$, which corresponds to the universe's age $t_{0}=13.76_{-0.14}^{+0.32}$ Gyrs. For the CC$+$Pantheon datasets, we find $t_{0}H_{0}=0.9697$, which corresponds to the age of the universe $t_{0}=13.82_{-0.11}^{+0.17}$ Gyrs. These estimates are in line with recent observations \cite{ref73,ref95,ref96,ref97}.

\section{Validity of the model}
In this section, we have performed some diagnostic tests to validate our derived model. These are as follows:
\subsection{Om diagnostic test}
We have explored the behavior of Om diagnostic function \cite{ref108} in solving the field equations in our derived model.  The Om diagnostic characteristic classifies the cosmic dark energy evolution of the expanding cosmos.  For a spatially homogenous universe, the Om diagnostic function is defined as
\begin{equation}\label{eq54}
	Om(z)=\frac{\left(\frac{H(z)}{H_{0}}\right)^{2}-1}{(1+z)^{3}-1},
\end{equation}
where $H(z)/H_{0}$ is the normalized Hubble function defined in Eq.\,\eqref{eq31}. The slope of the Om function categorize different dark energy model stages. For instance, the negative slope depicts the quintessence phase and positive slope represent the phantom scenarios of the model while the constant slope corresponds the $\Lambda$CDM behavior. For our model, the Om diagnostic function is derived by using the Eqs.\,\eqref{eq31} and \eqref{eq54} and is given below as:
\begin{equation}\label{eq55}
	Om(z)=\frac{\frac{\Omega_{m0}(1+z)^{3}+\Omega_{r0}(1+z)^{4}+\Omega_{\eta}}{\Omega_{m0}+\Omega_{r0}+\Omega_{\eta}+\frac{3(\beta+\gamma)H_{0}^{2}}{2\pi}\Omega_{m0}[1-(1+z)^{3}]+\frac{15\gamma\,H_{0}^{2}}{8\pi}\Omega_{r0}[1-(1+z)^{4}]}-1}{(1+z)^{3}-1}.
\end{equation} 
Figure 7 depicts the variation of Om diagnostic function over $z$ and one can observe that $Om(z)$ is an non-decreasing function of $z$ over the interval $[-1, 3]$ that reveals the phantom behavior of the model and also, one can see that at late-time $Om(z)$ function becomes a constant which indicates the tendency of the model to the $\Lambda$CDM stage at late time.
\begin{figure}[H]
	\centering
	\includegraphics[width=10cm,height=8cm,angle=0]{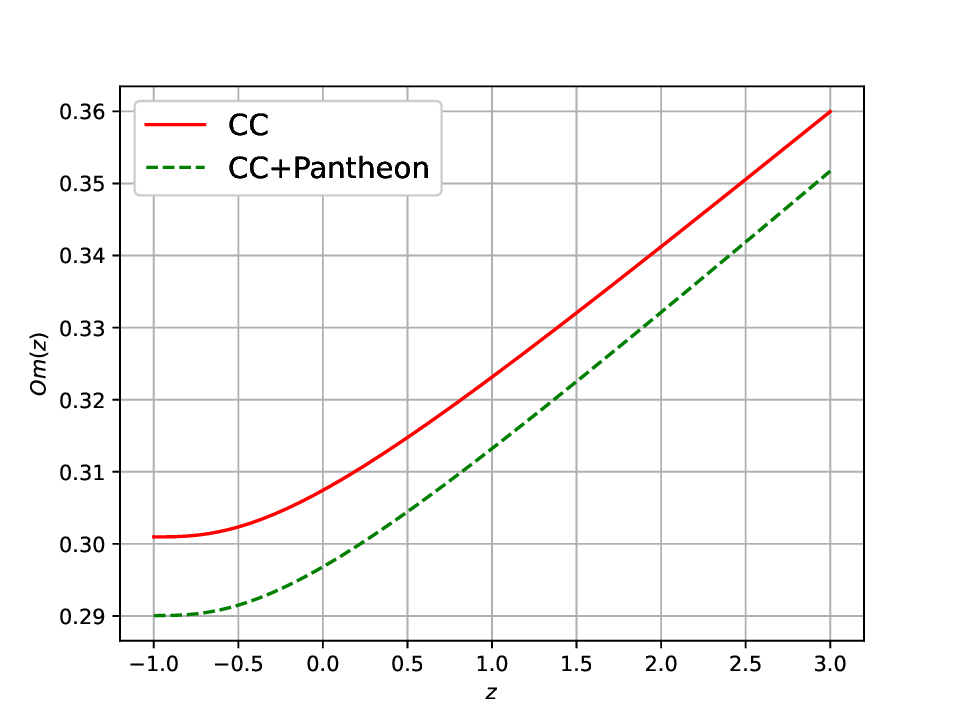}
	\caption{The variation of Om diagnostic parameter $Om(z)$ versus $z$.}
\end{figure}

\subsection{Causality test}

The physical acceptability and stability can be understand by investigating the behavior of squared sound speed $c_{s}^{2}$ (causality test). The metric perturbation using squared sound speed is defined as \cite{ref109}
\begin{equation}\label{eq56}
	c_{s}^{2}=c^{2}\frac{dp}{d\rho},
\end{equation}
where $c$ is the velocity of light and is taken as unity in cosmic unit. The positive value of $c_{s}^{2}/c^{2}$ indicates the stability of the model and its variation in the interval $[0, 1]$ confirms the physical acceptability of the model.\\
For the effective energy density and effective pressure, we derive the $c_{s}^{2}$ as below
\begin{equation}\label{eq57}
	c_{s}^{2}=-\frac{2}{3}+\frac{1}{3}(1+z)\left[\frac{H'}{H}+\frac{H''}{H'}\right],
\end{equation}
where
\begin{multline}\label{eq58}
	\frac{H'}{H}=\frac{1}{2}\left[\frac{3\Omega_{m0}(1+z)^{2}+4\Omega_{r0}(1+z)^{3}}{\Omega_{m0}(1+z)^{3}+\Omega_{r0}(1+z)^{4}+\Omega_{\eta}}\right.\\\left.+\frac{[9(\beta+\gamma)\Omega_{m0}(1+z)^{2}+15\gamma\Omega_{r0}(1+z)^{3}]H_{0}^{2}}{2\pi[\Omega_{m0}+\Omega_{r0}+\Omega_{\eta}+\frac{3(\beta+\gamma)H_{0}^{2}}{2\pi}\Omega_{m0}[1-(1+z)^{3}]+\frac{15\gamma\,H_{0}^{2}}{8\pi}\Omega_{r0}[1-(1+z)^{4}]]}\right],
\end{multline}
and
\begin{multline}\label{eq59}
	\frac{H''}{H}=\left(\frac{H'}{H}\right)^{2}+\frac{1}{2}\left[\frac{6\Omega_{m0}(1+z)+12\Omega_{r0}(1+z)^{2}}{\Omega_{m0}(1+z)^{3}+\Omega_{r0}(1+z)^{4}+\Omega_{\eta}}-\frac{[3\Omega_{m0}(1+z)^{2}+4\Omega_{r0}(1+z)^{3}]^{2}}{[\Omega_{m0}(1+z)^{3}+\Omega_{r0}(1+z)^{4}+\Omega_{\eta}]^{2}}\right.\\\left.+\frac{[18(\beta+\gamma)\Omega_{m0}(1+z)+45\gamma\Omega_{r0}(1+z)^{2}]H_{0}^{2}}{2\pi[\Omega_{m0}+\Omega_{r0}+\Omega_{\eta}+\frac{3(\beta+\gamma)H_{0}^{2}}{2\pi}\Omega_{m0}[1-(1+z)^{3}]+\frac{15\gamma\,H_{0}^{2}}{8\pi}\Omega_{r0}[1-(1+z)^{4}]]}\right.\\\left.+\frac{[9(\beta+\gamma)\Omega_{m0}(1+z)^{2}+15\gamma\Omega_{r0}(1+z)^{3}]^{2}H_{0}^{4}}{4\pi^{2}[\Omega_{m0}+\Omega_{r0}+\Omega_{\eta}+\frac{3(\beta+\gamma)H_{0}^{2}}{2\pi}\Omega_{m0}[1-(1+z)^{3}]+\frac{15\gamma\,H_{0}^{2}}{8\pi}\Omega_{r0}[1-(1+z)^{4}]]^{2}}\right].
\end{multline}
Figure 8 shows the variation of $c_{s}^{2}/c^{2}$ over the redshift $z$ as expressed by Eq.\,\eqref{eq57}. We measure the present value of $c_{s}^{2}/c^{2}=0.0280$ for CC data and $c_{s}^{2}/c^{2}=0.0303$ for CC+Pantheon datasets. Figure 8 shows that the function $c_{s}^{2}/c^{2}$ increases with $z$ and fluctuates as $0\leq\,c_{s}^{2}/c^{2}\leq1$ over the redshift interval $[-1, 3]$, confirming the physical acceptability and stability of our derived model.
\begin{figure}[H]
	\centering
	\includegraphics[width=10cm,height=8cm,angle=0]{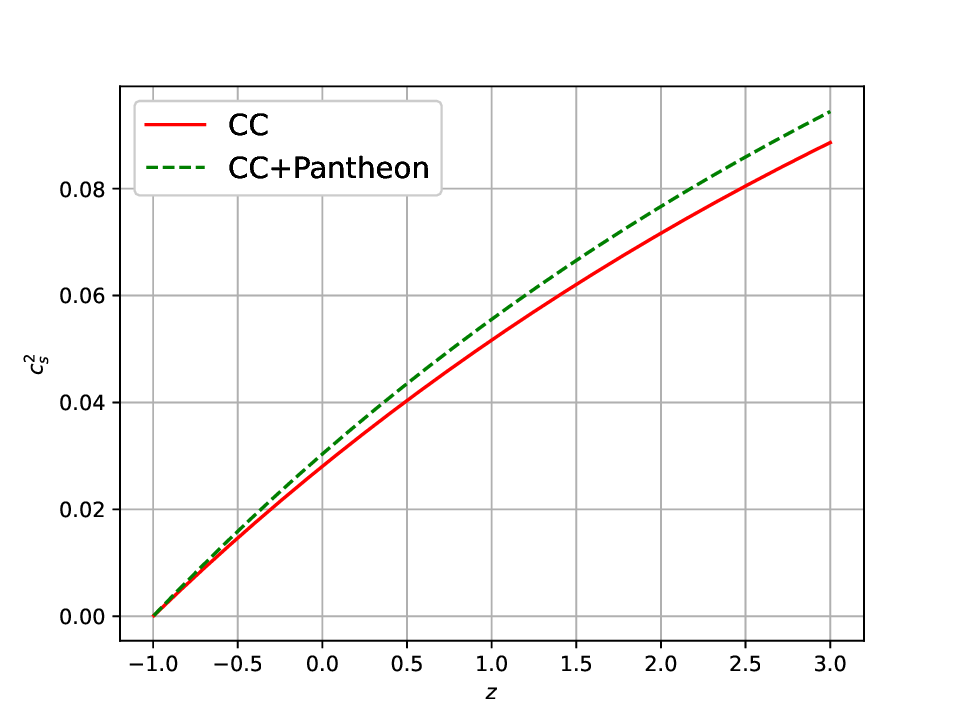}
	\caption{The variation of squared sound speed $c_{s}^{2}$ versus $z$.}
\end{figure}

\subsection{Energy conditions}

The physical consistency and different dark energy evolution stages of the model can be understood by the notion of energy conditions. The basic concepts of the energy conditions are presented through the Raychaudhuri equations, which imply that for the attractive nature of gravity, the energy density of the model must be positive \cite{ref110}. There are four energy conditions, namely, the null (NEC), weak (WEC), dominant (DEC) and strong (SEC) energy conditions as suggested in \cite{ref83, ref111,ref112}. For a flat, homogeneous and isotropic model, these energy conditions are defined as below.\\
\noindent{\bf NEC:} Null Energy Condition: $\rho_{eff} + p_{eff} \geq 0$,\\
{\bf WEC:} Weak Energy Conditions: $\rho_{eff} \geq 0$, $\rho_{eff} + p_{eff} \geq 0$,\\
{\bf DEC:} Dominant Energy Conditions: $\rho_{eff} \geq \lvert p_{eff} \rvert$ i.e. $\rho_{eff} \pm p_{eff} \geq 0$, \\
{\bf SEC:} Strong Energy Conditions: $\rho_{eff} + p_{eff} \geq 0$, $\rho_{eff} + 3p_{eff} \geq 0$.\\
where the effective energy density $\rho_{eff}$ and effective pressure $p_{eff}$ are defined in Eq.\,\eqref{eq44}.\\
The graphical representations of the energy conditions over redshift $z$ are shown in Figures 9a and 9b, respectively for CC datasets and CC+Pantheon datasets. From Figures 9a and 9b, one can observe that all the energy conditions are satisfied over the redshift $[-1, 3]$ except the strong energy conditions which violated for $z<z_{t}$. This violation of SEC causes the creation of exotic matter and consequently resultant of accelerating phase in the expansion of the universe.\\
\begin{figure}[H]
	\centering
	a.\includegraphics[width=8cm,height=7cm,angle=0]{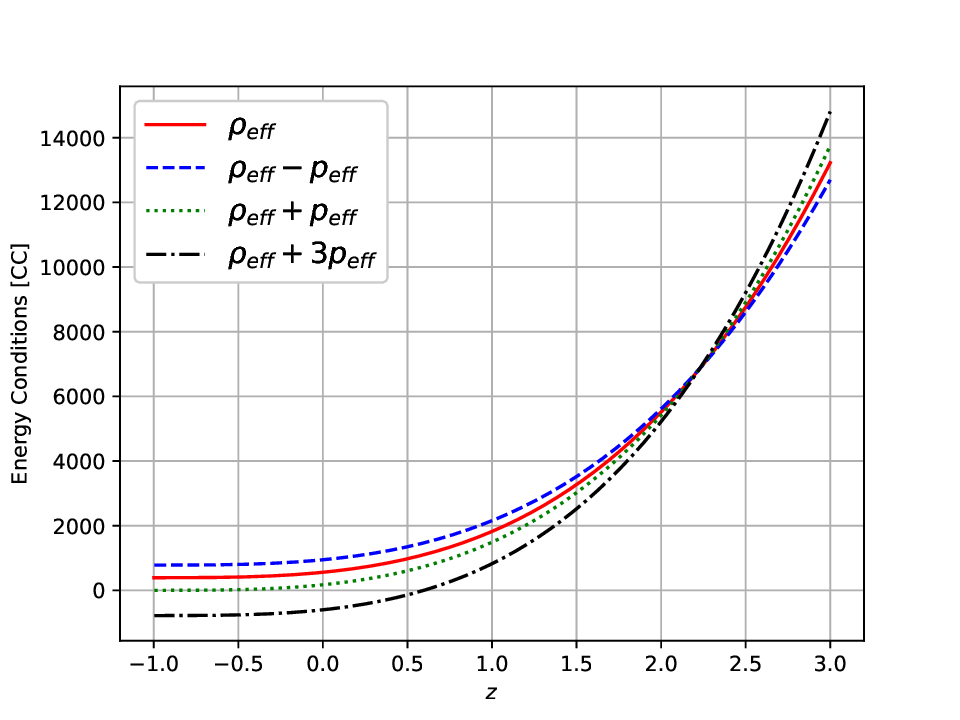}
	b.\includegraphics[width=8cm,height=7cm,angle=0]{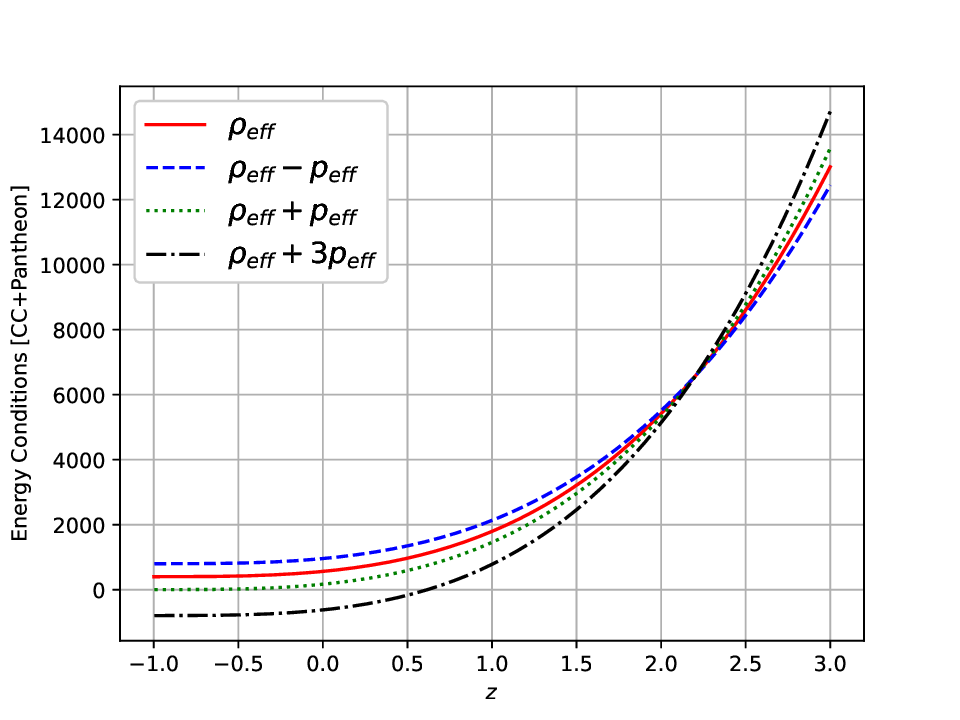}
	\caption{The variation of effective energy conditions versus $z$.}
\end{figure}

\begin{figure}[H]
	\centering
	a.\includegraphics[width=8cm,height=7cm,angle=0]{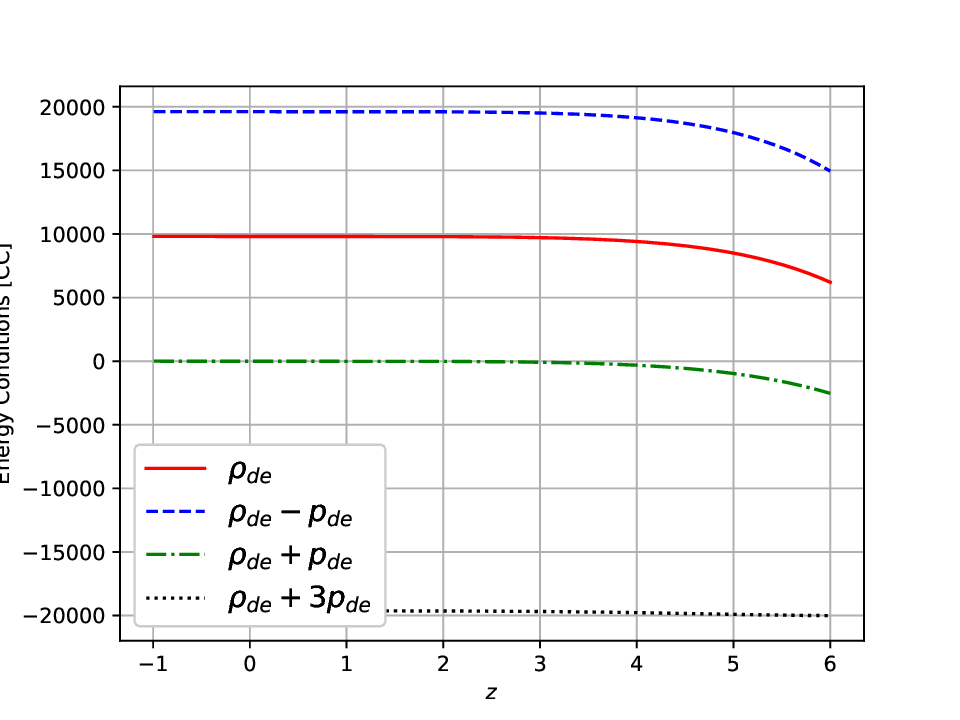}
	b.\includegraphics[width=8cm,height=7cm,angle=0]{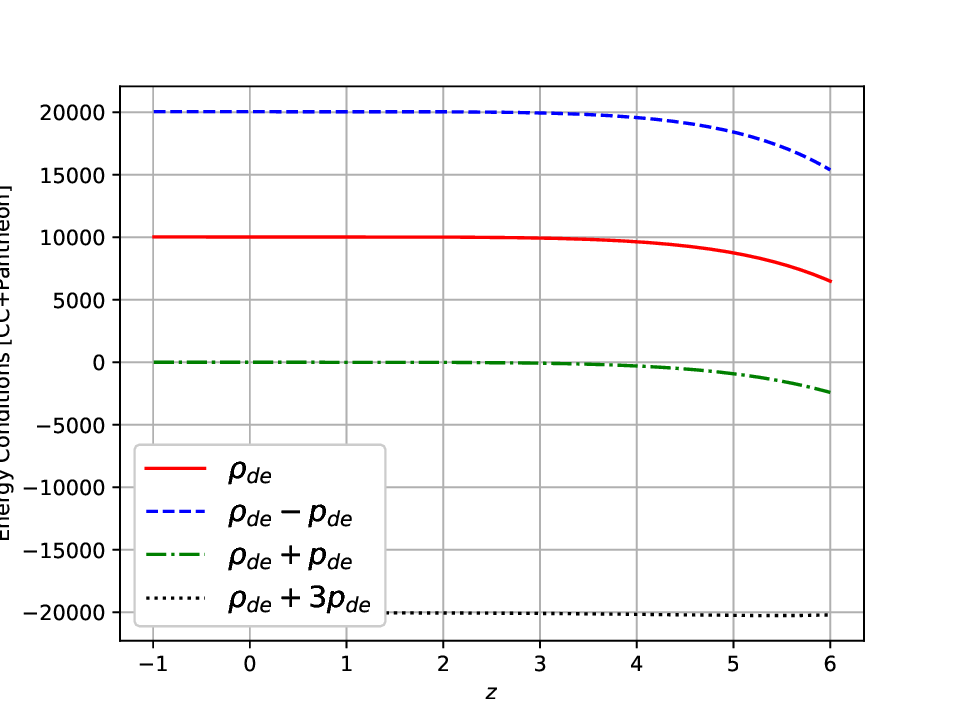}
	\caption{The variation of dark energy conditions versus $z$.}
\end{figure}
Figures 10a and 10b show the energy conditions for the dark sector as specified by $\rho_{de}$ and $p_{de}$ in the CC and CC+Pantheon datasets, respectively.  Figures 10a and 10b demonstrate that all of the energy criteria for the dark sector are violated, revealing the phantom dark energy scenario of the expanding cosmos.

\section{Conclusions}
%
In the present work, we have discussed dark energy models in $f(R,T,L_{m})$-gravity. We have solved the field equations for the Lagrangian function $f(R,T,L_{m})= \alpha\,R+\beta\,RT+\gamma\,RL_{m}-\eta$ in a flat, homogeneous and isotropic spacetime universe with $\alpha$, $\beta$, $\gamma$ and $\eta$ as coupling constants. We have derived the Hubble function $H(z)$ in terms of $H_{0}$, $\Omega_{m0}$, $\Omega_{r0}$, $\Omega_{\eta}$, $\beta$ and $\gamma$. We constrained the model parameters using the MCMC analysis of joint datasets cosmic chronometer and Pantheon samples to make our model consistent with the physically observed universe. Using these estimated values of model parameters, we have explored the cosmic evolution history of the universe through the investigation of deceleration parameter, effective equation of state, dark energy equation of state, total dark energy density parameters, age of the universe etc. We have also preformed the Om diagnostic test, causality test, tests for energy conditions for the physical acceptability and stability of the model.\\

The value of the Hubble constant has been measured as $H_{0}=67.9\pm3.1$ km/s/Mpc using CC data. In a joint analysis of CC and Pantheon datasets, the value found is $H_{0}=68.6\pm1.9$ km/s/Mpc.  We have quantified the values of dimensionless parameters.  The values obtained are $\Omega_{m0}=0.29\pm0.13, 0.266_{-0.091}^{+0.11}$, $\Omega_{r0}=0.020\pm0.011, 0.020_{-0.012}^{+0.013}$, and $\Omega_{\eta}=0.69_{-0.16}^{+0.26}, 0.70_{-0.14}^{+0.27}$, corresponding to the two datasets CC and CC+Pantheon.  The coupling constant $\beta=\gamma$ is measured as $\beta=-0.047\pm0.028\times10^{-7}$ for the CC dataset and $\beta=-0.052_{-0.039}^{+0.025}\times10^{-7}$ for the CC+Pantheon dataset.  The values of $\alpha$ have been measured as 1.03 and 0.986 for the CC and CC+Pantheon datasets, respectively.  The estimations align well with the observations made.  These measurements have been utilized in the analysis of various physical cosmological parameters.\\

We have discovered a transit universe that is currently accelerating and was decelerating in the past. We have found the present value of the deceleration parameter $q_{0}=-0.5388, -0.5547$ and measured the transition redshift $z_{t}=0.5965, 0.6177$, along two datasets, CC and CC+Pantheon, respectively, which are compatible with estimated values in \cite{ref98,ref99,ref100,ref101,ref102,ref103,ref104,ref105,ref106,ref107}. We have measured the present value of the effective EoS parameter $\omega_{eff}=-0.6925$ for CC and $\omega_{eff}=-0.7031$ for CC+Pantheon data, which are compatible with the effective EoS parameter value corresponding to $\Lambda$CDM. We have studied the behavior of the dark energy EoS parameter $\omega_{de}$, which depicts the phantom phase of the model. We have found the current value of $\omega_{de}\approx-1$. We have studied the variations of total energy density parameters $\Omega_{m}$ and $\Omega_{de}$ and found early matter dominated $(\Omega_{m}, \Omega_{de})\to(1, 0)$ as $z\to\infty$, while dark energy dominated at late-time universe $(\Omega_{m}, \Omega_{de})\to(0, 1)$ as $z\to-1$. We have found the current value of $\Omega_{m}\approx0.3$ and $\Omega_{de}\approx0.7$. Our Om diagnostic test of the model reveals the phantom evolution of the model. The behavior of the causality test of the model confirms that our universe model is physically viable, acceptable, and stable in nature. The study of energy conditions indicates a late-time accelerating phase with $\omega_{de}\leq-1$. We have measured the present age of the universe model as $t_{0}=13.76_{-0.14}^{+0.32}$ Gyrs for CC data, and for CC$+$Pantheon datasets, we found the age of the universe $t_{0}=13.82_{-0.11}^{+0.17}$ Gyrs.\\

The signs and values of coupling constants ($\alpha$, $\beta$, $\gamma$, and $\eta$) in $f(R, T, L_{m})=\alpha\,R+(\beta\,T+\gamma\,L_{m})R-\eta$ refer to, how different amounts of matter sources affect the behavior of the expanding universe. One can choose different signs of these constants to get different cosmic scenarios, which tend to the $\Lambda$CDM stage at late-time. Thus, utilizing a non-linear version of $f(R,T,L_{m})$ gravity theory, we have discovered a new type of phantom dark energy model that includes a transition phase, and in which the dark energy arises from the combined relationship between matter and geometry. The intriguing scenarios of dark energy models in $f(R,T,L_{m})$ gravity will encourage researchers to further explore this theory to reveal the history of the universe's evolution.

\section*{Acknowledgments}
The authors are thankful to IUCAA Center for Astronomy Research and Development (ICARD), CCASS, GLA University, Mathura, India, and United University, Prayagraj, India for providing facilities and support where part of this work is carried out.

\section*{Data Availability Statement}
No data is associated in the manuscript.



\end{document}